\documentclass[twocolumn,floatfix,preprintnumbers,citeautoscript,showpacs]{revtex4} 
\usepackage{graphicx}
\usepackage{amsmath,amssymb,graphics}
\usepackage{bbm} 
\usepackage{mathrsfs}	
\usepackage[T1]{fontenc}

\def\lsim{\mathrel{\raise.3ex\hbox{$<$\kern-.75em\lower1ex\hbox{$\sim$}}}}
\def\gsim{\mathrel{\raise.3ex\hbox{$>$\kern-.75em\lower1ex\hbox{$\sim$}}}}

\begin{document}
\preprint{DESY 07-158}

\title{Gamma Ray Spectrum from Gravitino Dark Matter Decay}

\author{Alejandro Ibarra}
\author{David Tran}

\affiliation{Deutsches Elektronen-Synchrotron DESY, Hamburg, Germany}

\pacs{95.35.+d, 11.30.Pb, 98.70.Rz}

\begin{abstract}
Gravitinos are very promising candidates for the cold dark matter
of the Universe. Interestingly, to achieve a sufficiently long
gravitino lifetime, $R$-parity conservation is not required, 
thus preventing any dangerous cosmological influence of the next-to-lightest
supersymmetric particle. When $R$-parity is violated, gravitinos
decay into photons and other particles with a lifetime much longer than the
age of the Universe, producing a diffuse gamma
ray flux with a characteristic spectrum that could be measured
in future experiments, like GLAST or AMS-02. 
In this letter we compute the energy spectrum 
of photons from gravitino decay and discuss its main qualitative features.

\end{abstract}

\maketitle

There is mounting evidence that dark matter is ubiquitous
in our Universe~\cite{Bertone:2004pz}.
Since the necessity of dark matter was realized, many 
particle physics candidates have been proposed. Among the most
interesting stands the gravitino~\cite{Pagels:1981ke}, 
the supersymmetric counterpart of the graviton, which arises 
when global supersymmetry is promoted to a local symmetry.
If the gravitino is the lightest supersymmetric particle (LSP), it
constitutes an excellent candidate for the cold dark
matter of the Universe. 
The thermal relic density of the gravitino
is calculable in terms of very few parameters, 
the result being~\cite{Bolz:2000fu}
\begin{equation}
\Omega_{3/2} h^2\simeq 0.27
    \left(\frac{T_R}{10^{10}\,\text{GeV}}\right)
    \left(\frac{100 \,\text{GeV}}{m_{3/2}}\right)
    \left(\frac{m_{\widetilde g}}{1\,\text{TeV}}\right)^2\;,
\label{relic-abundance}
\end{equation}
while the relic density inferred by WMAP for the $\Lambda$CDM model is
$\Omega_\text{CDM} h^2\simeq0.1$~\cite{Spergel:2006hy}.
In this formula, $T_R$ is the reheating temperature of the Universe,
$m_{3/2}$ is the gravitino mass and $m_{\widetilde g}$ is the 
gluino mass. It is indeed remarkable that the correct relic density
can be obtained for typical supersymmetric parameters, 
$m_{3/2}\sim 100\,\text{GeV}$, $m_{\widetilde g}\sim 1\,\text{TeV}$,
and a high reheating temperature, $T_R\sim 10^{10}\,\text{GeV}$, as
required by baryogenesis through the mechanism of 
thermal leptogenesis~\cite{Fukugita:1986hr}.

Whereas the gravitino as the 
LSP leads to a cosmology consistent with observations, 
the cosmology of the Next-to-Lightest Supersymmetric Particle (NLSP) 
is much more problematic. In supersymmetric model building, in order to 
prevent too rapid proton decay, it is common to invoke a discrete symmetry
called $R$-parity. When $R$-parity is exactly conserved, the NLSP can
only decay into gravitinos and Standard Model particles with a decay
rate strongly suppressed by the Planck mass. As a result, the NLSP is 
typically present in the Universe at the time of Big Bang 
nucleosynthesis, jeopardizing the successful predictions of the
standard nucleosynthesis scenario. In most supersymmetric scenarios,
the NLSP is either a neutralino or a stau. On one hand, if the
NLSP is a neutralino, its late decay into hadrons can
dissociate the primordial elements~\cite{Kawasaki:2004qu}. 
On the other hand, if the NLSP is a stau, it can form a 
bound state with $^4$He, catalyzing the production of 
$^6$Li~\cite{Pospelov:2006sc}. As a result, the abundance 
of $^6$Li is increased by a factor 300--600, in stark 
conflict with observations~\cite{Hamaguchi:2007mp}.

Several scenarios have been proposed that circumvent
the above-mentioned difficulties~\cite{solutions}. The simplest,
albeit the most radical one, is based on the assumption that
$R$-parity is not exactly conserved~\cite{Buchmuller:2007ui}.
In fact, although experiments set very stringent bounds
on $R$-parity violation, there is no deep theoretical reason why 
it should be exactly conserved. If $R$-parity is mildly violated,
the NLSP decays into Standard Model particles well before 
the first nucleosynthesis reactions take place, thus not posing 
a jeopardy for the Standard Model predictions. Remarkably, even though 
the gravitino is no longer stable when $R$-parity is violated, it
still constitutes a viable dark matter candidate~\cite{Takayama:2000uz}. 
To be precise, a consistent thermal history of the Universe with
gravitino dark matter, thermal leptogenesis and successful
primordial nucleosynthesis requires the lepton number violating
Yukawa couplings to lie between $10^{-14}$ and $10^{-7}$
\cite{Buchmuller:2007ui}, which translates into gravitino lifetimes
in the range $10^{23}-10^{37}$ s for $m_{3/2}\sim 100$ GeV,
which are much longer than the age of the Universe.
Nevertheless, gravitino decays could be happening at a sufficiently
high rate for the decay products to be detectable in future experiments.

In this letter we will concentrate on the photons produced
in gravitino decays, although in general other stable particles 
are also produced, such as electrons, protons, neutrinos
and their antiparticles. Demanding a high reheating temperature 
for the Universe as suggested by thermal leptogenesis, 
$T_R\gsim 10^9$ GeV~\cite{bound}, 
it follows from Eq.~(\ref{relic-abundance}) that 
the gravitino mass has to be $m_{3/2}\gsim 5$ GeV for 
typical gluino massses. Consequently, we expect the photons
from gravitino dark matter decay in the energy range of a 
few GeV, {\it i.e.} in the gamma ray energy range. 

The Energetic Gamma Ray Experiment Telescope (EGRET) aboard
the Compton Gamma Ray Observatory measured gamma rays
in the energy range between 30 MeV to 100 GeV. After 
subtracting the galactic foreground emission, the residual
flux was found to be roughly isotropic and thus attributed to 
extragalactic sources. The first analysis of the EGRET data
by Sreekumar {\it et al.}~\cite{egret} gave an extragalactic flux 
with an energy spectrum described by the power law
\begin{equation}
E^2 \frac{dJ}{dE} = 1.37 \times 10^{-6}\ 
\left(\frac{E}{1~\text{GeV}}\right)^{-0.1} 
(\text{cm}^2 \text{str}~\text{s})^{-1}\text{GeV}  
\end{equation}
in the energy range 50 MeV--10 GeV. The improved analysis
of the galactic foreground by Strong {\it et al.}~\cite{smr05},
optimized in order to reproduce the galactic emission,
shows a power law behavior between  50 MeV--2 GeV, but
a clear excess between 2--10 GeV, roughly the same
energy range where one would expect a signal from gravitino
decay. Although it is very tempting to look for explanations
for this excess in terms of gravitino decays, in view of all
the systematic uncertainties involved in the extraction of the
signal from the galactic foreground, we will not attempt to fit
our predicted flux to the EGRET data. Nonetheless, we will 
show later the EGRET data  superimposed with our predicted flux 
for comparison.

The total gamma ray flux received from gravitino dark matter
decay receives two main contributions. The first
one stems from the decay of gravitinos at cosmological
distances, giving rise to a perfectly isotropic extragalactic 
diffuse gamma ray background. Defining $dN_\gamma/dE$ as the
gamma ray spectrum produced in the gravitino decay, the flux 
received at the Earth with extragalactic origin
has the following expression:
\begin{equation}
\left[E^2 \frac{dJ}{dE}\right]_{\text{eg}} =
 \frac{2 E^2}{m_{3/2}} C_\gamma \int_1^\infty dy 
\frac{dN_\gamma}{d(E y)} \frac{y^{-3/2}}
{\sqrt{1 + \Omega_\Lambda/\Omega_M y^{-3}}}\;,
\label{extgal-flux}
\end{equation}
where $y=1+z$, $z$ being the redshift, and
\begin{equation}
C_\gamma = \frac{\Omega_{3/2} \rho_c}{8 \pi \tau_{3/2} H_0 \Omega_M^{1/2}}
\simeq 10^{-7}\ (\text{cm}^2\text{s}~\text{str})^{-1} \text{GeV}
\left(\frac{\tau_{3/2}}{ 10^{28}\ \mbox{s}} \right)^{-1}\,.
\end{equation}
Here, $ \Omega_{3/2} $, $ \Omega_M $ and $ \Omega_{\Lambda} $ are the 
gravitino, matter and cosmological constant density parameters, 
respectively, $ \rho_c $ is the critical 
density, $ \tau_{3/2} $ the gravitino lifetime, and $ H_0 $
the present value of the Hubble parameter.

In addition to the cosmological contribution, the total gamma 
ray flux also receives a contribution from the decay of
gravitinos in the Milky Way halo. This contribution reads:
\begin{equation}
\left[E^2 \frac{dJ}{dE}\right]_{\text{halo}} = 
\frac{2 E^2}{m_{3/2}} \frac{dN_\gamma}{dE} 
\frac{1}{8 \pi \tau_{3/2}} \int_\text{los} \rho_\text{halo}(\vec{l}) d\vec{l}
\;,
\label{halo-flux}
\end{equation}
The integration extends over the line of sight, so the halo contribution
has an angular dependence on the direction of observation, yielding 
an anisotropic gamma ray flux that in EGRET could resemble
an isotropic extragalactic flux.
Namely, in the energy range 0.1-10 GeV, 
the anisotropy between the Inner Galaxy region
($|b|>10^\circ, 270^\circ\leq l \leq 90^\circ$)
and the Outer Galaxy region 
($|b|>10^\circ, 90^\circ\leq l \leq 270^\circ$)
is just a 6\%, well compatible with the EGRET data~\cite{smr05}
(for a detailed discussion applied to GLAST, see \cite{Bertone:2007aw}). 
In Eq.~(\ref{halo-flux}), $\rho_\text{halo}$ 
stands for the dark matter distribution in the Milky Way halo. 
For our numerical analysis we will adopt a Navarro-Frenk-White 
density profile~\cite{Navarro:1996gj}
\begin{equation}
\rho_\text{halo}(r)\simeq \frac{\rho_h}{r/r_c (1+r/r_c)^2}\;,
\end{equation}
where $r$ is the distance to the Galactic center,
$r_c\simeq20\,\text{kpc}$ is the critical radius 
and $\rho_h\simeq0.33\,\text{GeV}\,\text{cm}^{-3}$.

In Eqs.~(\ref{extgal-flux},\ref{halo-flux}) the only undetermined
quantity is the energy spectrum of photons produced in the gravitino decay, 
$dN_\gamma/dE$, which depends crucially on the gravitino
mass. If the gravitino is lighter than the $W^\pm$ bosons,
it decays mainly into a photon and a neutrino by means
of the photino-neutrino mixing that arises when
$R$-parity is violated~\cite{Takayama:2000uz}.
Therefore, the spectrum is simply
\begin{equation}
\frac{dN_\gamma}{dE}\simeq\delta\left(E-\frac{m_{3/2}}{2}\right)\;.
\end{equation}
For this case, it was found in~\cite{Buchmuller:2007ui,Bertone:2007aw}
that the total gamma ray flux received is dominated by the
monochromatic line from the decay of gravitinos in 
our Milky Way halo, while the redshifted line from the decay 
of gravitinos at cosmological distances is somewhat fainter. 

On the other hand, if the gravitino is heavier than
the $W^\pm$ or $Z^0$ bosons, new decay modes are open.
In addition to the decay mode into a photon and a neutrino that 
follows from the photino-neutrino mixing, $U_{\tilde \gamma \nu}$,
the gravitino can also decay into a $W^\pm$ boson and a charged lepton,
through the mixing charged wino-charged lepton, $U_{\tilde W \ell}$,
or into a $Z^0$ boson and a neutrino, through the mixing 
zino-neutrino, $U_{\tilde Z\nu}$. The decay rates can be 
straightforwardly computed from the interaction Lagrangian 
of a gravitino with a gauge boson and a 
fermion~\cite{Cremmer:1982en}. Neglecting the masses of the 
final fermions, the result for each decay mode can
be approximated by:
\begin{eqnarray}
\Gamma(\psi_{3/2}\rightarrow \gamma \nu)\simeq
\frac{1}{32\pi}|U_{\tilde \gamma\nu}|^2\frac{m^3_{3/2}}{M_P^2}\;,\nonumber \\
\Gamma(\psi_{3/2}\rightarrow W^\pm \ell^\mp)\simeq
\frac{1}{16\pi}|U_{\tilde W \ell}|^2\frac{m^3_{3/2}}{M_P^2}
f\left(\frac{M_W}{m_{3/2}}\right)\;,\nonumber \\
\Gamma(\psi_{3/2}\rightarrow Z^0 \nu)\simeq
\frac{1}{32\pi}|U_{\tilde Z\nu}|^2\frac{m^3_{3/2}}{M_P^2}
f\left(\frac{M_Z}{m_{3/2}}\right)\;,
\label{decayrates}
\end{eqnarray}
where $f(x)=1-\frac{4}{3}x^2+\frac{1}{3}x^8$.

The fragmentation of the $W^\pm$ and the $Z^0$ gauge bosons will 
eventually produce photons, mainly from the decay of neutral
pions. We have simulated the fragmentation of the 
gauge bosons with the event generator
PYTHIA 6.4~\cite{Sjostrand:2006za} and calculated the spectra of photons
in the $W^\pm$ and $Z^0$ channels, which we denote by 
$dN_\gamma^W/dE$ and $dN_\gamma^Z/dE$, respectively.
The total spectrum is  given by:
\begin{eqnarray}
\frac{dN_\gamma}{dE}\simeq\text{BR}(\psi_{3/2}\rightarrow \gamma \nu)
\delta\left(E-\frac{m_{3/2}}{2}\right)+\nonumber \\
\text{BR}(\psi_{3/2}\rightarrow W \ell)\frac{dN^W_\gamma}{dE}+
\text{BR}(\psi_{3/2}\rightarrow Z^0 \nu)\frac{dN^Z_\gamma}{dE} \,. 
\label{inj-spectrum}
\end{eqnarray}

The branching ratios in the different decay channels are 
determined by the size of the $R$-parity breaking mixing
parameters, $U_{\tilde \gamma\nu}$, $U_{\tilde Z\nu}$ and
$U_{\tilde W \ell}$, and by the kinematical function $f(x)$
defined after Eq.~(\ref{decayrates}). The mixing parameters 
stem from the diagonalization of the $7\times 7$
neutralino-neutrino and $5\times 5$ chargino-charged
lepton mass matrices, whose explicit form can 
be found in the vast existing literature on $R$-parity 
violation~\cite{Barbier:2004ez}.
The precise expression for the mixing parameters in terms of 
the $R$-parity breaking couplings in the Lagrangian is fairly
cumbersome and will not be reproduced here. However, to derive 
the branching ratios, only the ratio among them is relevant, and not
their overall value. 

To derive the relation between $U_{\tilde \gamma\nu}$ and 
$U_{\tilde Z\nu}$, we first note that the photino does not
couple directly to the neutrino (since neutrinos do not couple
to photons). Nevertheless, an effective photino-neutrino
mixing is generated through the mixing photino-zino
and the mixing zino-neutrino.
The result reads:
\begin{equation}
|U_{\widetilde \gamma \nu}|\simeq 
\left|\frac{\text{M}^n_{\widetilde\gamma \widetilde Z}}
{\text{M}^n_{\widetilde \gamma \widetilde \gamma}} \right|
|U_{\widetilde Z \nu}|\;.
\end{equation}
Therefore, the relation between $U_{\tilde \gamma\nu}$ and 
$U_{\tilde Z\nu}$ follows from the $2\times 2$ gaugino 
sub-block of the neutralino mass matrix, that 
in the $(-i\widetilde{\gamma}, -i {\widetilde Z})$ basis reads
\begin{equation} 
\text{M}^n_{2\times 2} =\left(\begin{matrix}
M_1 c^2_W + M_2 s^2_W & (M_2-M_1) s_W c_W  \\
(M_2-M_1) s_W c_W & M_1 s^2_W + M_2 c_W^2 
\end{matrix}\right)\;.
\label{neutralino-mass}
\end{equation}
Here, $M_1$ and $M_2$ are the $U(1)_Y$ and $SU(2)_L$ gaugino masses, 
and $c_W$ ($s_W$) denotes the cosine (sine) of the weak mixing angle.
Therefore, 
\begin{equation}
|U_{\widetilde \gamma \nu}|\simeq 
\left[\frac{(M_2-M_1) s_W c_W }{ M_1 c^2_W + M_2 s_W^2}\right]
|U_{\widetilde Z \nu}|\;,
\end{equation}
which depends only on the gaugino masses at low energies.
Assuming gaugino mass universality at the Grand Unified Scale,
$M_X=2\times 10^{16}\,\text{GeV}$, we obtain at 
the electroweak scale $M_2/M_1\simeq 1.9$, which yields
$|U_{\widetilde \gamma \nu}|\simeq 0.31 |U_{\widetilde Z \nu}|$.

The mixing parameter $U_{\widetilde W \ell}$,
on the other hand, is related to $U_{\widetilde Z \nu}$
by $SU(2)_L$ gauge invariance. The relation approximately reads:
\begin{equation}
|U_{\widetilde W \ell}|\simeq \sqrt{2} c_W
\left|\frac{\text{M}^n_{\widetilde Z\widetilde Z}}
{M_{\widetilde W}}\right||U_{\widetilde Z \nu}|\;,
\label{UWtau1}
\end{equation}
where $M_{\widetilde W}=M_2$ is the wino mass at the electroweak
scale. Using Eq.~(\ref{neutralino-mass}), we finally obtain
\begin{equation}
|U_{\widetilde W \ell}|\simeq
\sqrt{2}c_W\frac{M_1 s^2_W + M_2 c_W^2}{M_2}|U_{\widetilde Z \nu}| \;,
\label{UWtau2}
\end{equation}
which under the assumption of gaugino mass universality at $M_X$
yields $|U_{\widetilde W \ell}|\simeq 1.09 |U_{\widetilde Z \nu}|$.
Hence, under this assumption, the mixing parameters
are in the ratio
\begin{equation}
|U_{\widetilde \gamma \nu}|:|U_{\widetilde Z \nu}|:|U_{\widetilde W \ell}|
\simeq 1:3.2:3.5\;,
\end{equation}
and thus the branching ratios for the different decay modes only
depend on the gravitino mass (see Table~\ref{tab:BRs}).

\begin{table}[t]
\caption{\label{tab:BRs}Branching ratios for gravitino decay
in different $R$-parity violating channels for different gravitino masses.}
\begin{ruledtabular}
\begin{tabular}{cccc}
$m_{3/2}$ & $\text{BR}(\psi_{3/2}\rightarrow \gamma \nu)$
&$\text{BR}(\psi_{3/2}\rightarrow W \ell)$ &
$\text{BR}(\psi_{3/2}\rightarrow Z^0 \nu)$\\
\hline
10 GeV & 1    & 0    & 0    \\
85 GeV & 0.66 & 0.34 & 0    \\
100 GeV& 0.16 & 0.76 & 0.08 \\
150 GeV& 0.05 & 0.71 & 0.24 \\
250 GeV& 0.03 & 0.69 & 0.28 
\end{tabular}
\end{ruledtabular}
\end{table}

Once the spectrum of photons from gravitino decay
has been computed, Eq.~(\ref{inj-spectrum}), it is 
straightforward to calculate the gamma-ray flux received 
at the Earth from our local halo and from cosmological distances, 
by using Eqs.~(\ref{extgal-flux},\ref{halo-flux}).
Assuming universality of gaugino masses at high energies, 
the photon flux received from gravitino decay depends 
essentially on the gravitino mass, which determines the 
shape of the energy spectrum, and the gravitino lifetime, 
which determines its overall normalization.

\begin{figure}[t]
\centerline{\includegraphics[width=8cm]{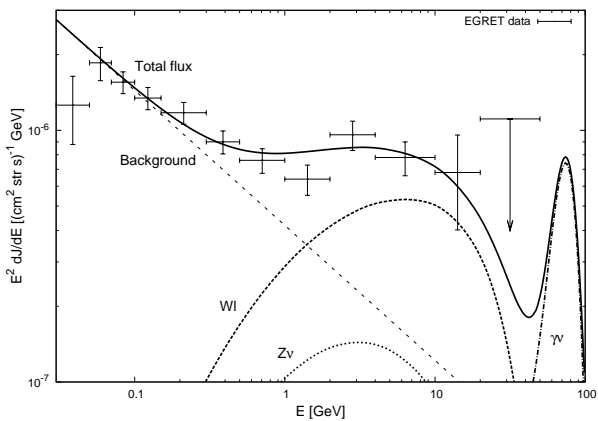}}
\caption{Contributions to the total gamma ray flux
for $m_{3/2}=150\,\text{GeV}$ and $\tau_{3/2}\simeq 
2\times 10^{26}\text{s}$ compared to the EGRET data.
In dotted lines we show
the photon flux from the fragmentation of the $Z$ boson, in
dashed lines from the fragmentation of the $W$ boson,
and in dot-dashed lines from the two body decay 
$\psi_{3/2}\rightarrow \gamma \nu$. The background is shown
as a long dashed line, while the total flux received is shown as a
thick solid line. 
\label{fig:flux}}
\end{figure}

In Fig.~\ref{fig:flux} we show the different contributions
to the gamma ray flux for $m_{3/2}=150\,\text{GeV}$ and 
$\tau_{3/2}\simeq 2\times 10^{26}\text{s}$.  To compare our results with 
the EGRET data~\cite{smr05}, also shown in the figure, we have averaged
the halo signal over the whole sky excluding a band of $\pm 10^\circ$
around the Galactic disk, and we have used an energy resolution 
of 15\%, as quoted by the EGRET collaboration in this energy range. 
The energy resolution of the detector is particularly
important to determine the width and the height of the monochromatic
line stemming from the two body decay $\psi_{3/2}\rightarrow \gamma \nu$.
The three contributions are dominated
by the halo component, the extragalactic component being
smaller by a factor of 2--3. Finally, we have adopted an 
energy spectrum for the extragalactic background 
described by the power law
$\left[E^2 \frac{dJ}{dE}\right]_\text{bg}=4\times 10^{-7}
\left(\frac{E}{\rm GeV}\right)^{-0.5}
(\text{cm}^2 \text{str}~\text{s})^{-1}\text{GeV}$,
in order to provide a qualitatively good agreement of the total
flux received with the data.

The predicted energy spectrum shows two qualitatively
different features. 
At energies between 1--10 GeV, we expect a continuous spectrum
of photons coming from the fragmentation of the gauge bosons.
As a result, the predicted spectrum shows a departure from the 
power law in this energy range that might be part of the 
apparent excess inferred from the EGRET data
by Strong {\it et al.}~\cite{smr05}.
The  upcoming space-based gamma ray experiments GLAST and AMS-02
will measure the energy spectrum with unprecedented accuracy,
providing very valuable information for the scenario
of decaying gravitino dark matter.

In addition to the continuous component, the energy spectrum
shows a relatively intense monochromatic line at 
higher energies arising from the decay channel
$\psi_{3/2}\rightarrow \gamma \nu$. This line could be observed
not only by GLAST or AMS-02 in the diffuse gamma background, but also
by ground-based Cherenkov telescopes such as MAGIC (with an energy 
threshold of 70 GeV) or VERITAS (50 GeV) in galaxies 
such as M31~\cite{Bertone:2007aw}.

The intense gamma line is very characteristic of this scenario, 
and the observation of this feature with the right intensity would 
support the gravitino dark matter decay hypothesis. 
While scenarios with neutralino dark matter also predict 
a continuous spectrum and a monochromatic line 
coming from the annihilation channels 
$\chi^0 \chi^0\rightarrow \gamma\gamma, Z \gamma$~\cite{Rudaz:1990rt},
these channels only arise at one loop
level, and thus the intensity of the monochromatic line is
greatly suppressed compared to the continuum. 
One should note, however, that the presence of an intense 
gamma line is not unique to the scenario with decaying gravitino 
dark matter and is also expected, for example, from the annihilation
of inert Higgs dark matter~\cite{Gustafsson:2007pc}.

To summarize, in this letter we have computed the gamma
ray flux from gravitino dark matter decay in scenarios
with $R$-parity violation. These scenarios are very appealing
theoretically, as they naturally lead to a 
history of the Universe consistent with thermal leptogenesis
and primordial nucleosynthesis. The predicted flux essentially
depends on two parameters: the gravitino mass, which determines the 
shape of the energy spectrum, and the gravitino lifetime, which determines 
its overall normalization. If the gravitino is lighter than
the $W^\pm$ and $Z^0$ gauge bosons, the predicted energy spectrum
is essentially monochromatic. On the other hand, if it is heavier, 
the energy spectrum consists of a continuous component and a 
relatively intense gamma ray line. This gamma ray flux might 
have already been observed by EGRET. Future experiments, such as
GLAST, AMS-02 or Cherenkov telescopes, will provide unique
opportunities to test the decaying gravitino dark matter scenario.

\smallskip
{\it Acknowledgements:} We are grateful to W. Buchm\"uller,
G. Bertone, J. Cortina, L. Covi, L. Pieri and F.D. Steffen 
for useful discussions and suggestions.


\end{document}